\documentclass[10pt]{iopart}
\usepackage{iopams}
\usepackage{graphicx}
\begin{document}

\title{Dynamics of Relativistic Solitons}

\author{Daniela Farina\dag\ and Sergei V. Bulanov\ddag \S}

\address{\dag\ Istituto di Fisica del Plasma ``Piero Caldirola'',
Consiglio Nazionale delle Ricerche, EURATOM-ENEA-CNR Association,
Milan, Italy}

\address{\ddag\ Advanced Photon Research Center, Kansai Research
Establishment, JAERI, Kizu, Kyoto 619-0215, Japan }

\address{\S A. M. Prokhorov Institute of General Physics,
 Russ. Acad. Sci., Moscow, Russia}

\begin{abstract}
Relativistic solitons are self-trapped, finite size, electromagnetic
waves of relativistic intensity that propagate  without diffraction
spreading. They have been predicted theoretically within the
relativistic fluid approximation, and  have been observed in
multi--dimensional particle in cell simulations of laser pulse
interaction with the plasma.  Solitons were observed in the
laser irradiated plasmas with the proton imaging technique as well.
This paper reviews many theoretical results on relativistic solitons
in electron--ion plasmas.
\end{abstract}

\section{Introduction}

Relativistic solitons are self--trapped, finite size,
electromagnetic waves of relativistic intensity $(eE/m \omega c \ge
1)$ that propagate without diffraction spreading. Many different
physical effects play a role in the formation of relativistic
solitons: dispersion effects due to the finite particle inertia,
nonlinearities due to relativistic mass increase, as well as
ponderomotive effects which force the plasma density redistribution.

The theoretical investigation of relativistic solitons in
electron--ion plasmas  is a relatively old problem in plasma
physics, which has been treated by many authors  in the past, and
has recently gained new attention in the literature
\cite{GT,MT,Lai,TT,KLS,KSK,KZ, Dimant,Esirkepov:98,prl01,ppr01,aipproc,poorn,mau,poorn:warm,jovo}.
The analyses have been performed mainly in  the framework of the 1D
relativistic fluid approximation, in which  solitons  are described
by the solutions of a set of coupled nonlinear differential
equations for the electrostatic and electromagnetic potentials with
suitable boundary conditions.

Relativistic solitons have been seen in   multi-dimensional particle
in cell (PIC) as well as in fluid simulations of laser pulse
interaction with the plasma \cite{sim92,sim95,sim99,Sentoku-99,
postsol,sim02,sim04,tush}. These solitons consist of electron
density depressions and intense electromagnetic field concentrations
with a larger amplitude and a lower frequency than those of the
laser pulse. Since a significant portion of the overall
electromagnetic energy is trapped in the form of solitons, solitary
waves can play an important role in the laser--plasma interaction.

\section{1D relativistic solitons}

Here, the theory of 1D circularly polarized solitons is presented
within the relativistic hydrodynamic approximation  used to describe
both the electron and ion components \cite{KLS,prl01,ppr01}. The plasma is assumed to be
cold with zero ion and electron temperature. Length, time, velocity,
momentum, vector and scalar potential, and density are normalized
over $c/\omega_{pe}$, $\omega_{pe}$, $c$, $m_{\alpha}c$,
$m_{e}c^{2}/e$, and $n_{0}$, respectively, being $\omega_{pe}=({4\pi
n_{0}e^{2}/m_{e}})^{1/2}$ the electron plasma frequency,
$m_{\alpha}$ the rest mass with $\alpha=e,i$, and $n_{0}$ the
unperturbed electron (and ion) density.  In the Coulomb gauge, the
Maxwell's equations for the vector and scalar potentials, ${\mathbf
A}$ and $\phi$, and the hydrodynamic equations (continuity, and
momentum
balance) for the densities $n_{\alpha}$ and the canonical momentum
${\mathbf P}_{\alpha}$ of electrons and ions can be written as
\begin{eqnarray}
& &\triangle {\mathbf A}-\frac{\partial^{2}}{\partial t^{2}} {\mathbf A} -\frac{\partial}{\partial t} \nabla \phi =  n_{e}{\mathbf v} _{e}-n_{i}{\mathbf v}_{i} \, , \label{eqa} \\
& &\triangle \phi =n_{e}-n_{i} \, , \label{eqn} \\ & &\frac{\partial}{\partial t} n_{\alpha}+\nabla \cdot (n_{\alpha} {\mathbf v} _{\alpha})=0, \label{eqcn} \\
& &\frac{\partial}{\partial t} {\mathbf P}_{\alpha} = -\nabla(\rho_{\alpha} \phi+ \gamma_{\alpha})+ {\mathbf v} _{\alpha}\times \nabla \times {\mathbf P}_{\alpha} \, , \label{eqp}
\end{eqnarray}
where ${\mathbf P}_{\alpha}$, and $\gamma_{\alpha}$ are related to the kinetic momentum
${\mathbf p}_{\alpha}$ by ${\mathbf P}_{\alpha}={\mathbf p}_{\alpha}+ \rho_{\alpha}{\mathbf A}$,
and $\gamma_{\alpha}=(1+|{\mathbf p}_{\alpha}|^{2})^{1/2}$
with the parameter $\rho_{\alpha}= (q_{\alpha}/q_{e})(m_{\alpha}/m_{e})$,
and ${\mathbf v}_{\alpha}={\mathbf p}_{\alpha}/\gamma_{\alpha}$ is the fluid velocity.

For the 1D case in which $\partial_{y}=\partial_{z}=0$, the vector
potential is assumed of the form (circular polarized wave)
\begin{equation}
A_{y}+iA_{z}=a(\xi )\exp {[i \theta(\xi)]} \, \exp {(-i\omega t +i k x )} \label{aperp}
\end{equation}
with $\xi=x-Vt$,  while all the other quantities,
$\phi $, $n_{\alpha }$, $\gamma _{\alpha }$,
and $p_{x \alpha }$, are assumed to depend only on the variable $\xi $.
In the present case, the relations $A_{x}=0$  and ${\bf P}_{\perp}=0$ are satisfied,
where the symbol $\perp$ refers to the direction of the laser pulse propagation.

Imposing as boundary condition at the point $\xi=-\infty$, $a=\pm
a_{0}$, $\phi =0$, $n_{\alpha}=1$, and $p_{x \alpha}=0$ (plasma at
rest), the longitudinal component of the kinetic momentum, the
energy, and the density of each species can be expressed as a
function of the potentials as $p_{x \alpha }=(V\psi _{\alpha
}-R_{\alpha })/(1-V^{2})$, $ \gamma _{\alpha }=(\psi _{\alpha
}-VR_{\alpha })/(1-V^{2}) $, $n_{\alpha }=V(\psi _{\alpha
}/R_{\alpha }-V)/(1-V^{2})$, where
$\psi_{\alpha}=\Gamma_{0\alpha}+\rho_{\alpha}\phi $,
$R_{\alpha}=[\psi _{\alpha}^{2}-(1-V^{2})\Gamma_{\alpha}^{2}]^{1/2}
$, $\Gamma_{\alpha}=\sqrt{1+\rho_{\alpha}^{2}a^{2}}$ and
$\Gamma_{0\alpha}=\sqrt{1+\rho_{\alpha}^{2}a_{0}^{2}}$.
Then, for an electron--ion plasma the following closed system of equations
for the potentials is obtained
\begin{eqnarray}
& &\frac{d^{2}\phi }{d\xi^{2} }=\frac{V}{1-V^2} \left( \frac{\psi _{e}}{R_{e}}-\frac{\psi _{i}}{R_{i}}\right) , \label{eq:phi} \\
& &\frac{d^{2}a }{d\xi^{2}}+ a\left( {\bar \omega}^{2}-{\bar k}^{2} \frac{a_{0}^{4}}{a^{4}} \right )= a\frac{V}{1-V^2}\left( \frac{1}{R_{e}}+ \frac{\rho }{R_{i}}\right) . \label{eq:a}\\
& &\frac{d \theta}{d \xi}=-{\bar k}\left ( 1- \frac{a_{0}^{2}}{a^{2}} \right ) \label{eq:th}
\end{eqnarray}
where ${\bar \omega}=(\omega-kV)/(1-V^2)$, ${\bar k}=(k-\omega V)/(1-V^2)$, and $\rho \equiv |\rho_{i}|=m_{e}/m_{i}$.

The system of equations (\ref{eq:phi},\ref{eq:a}) describes coupled
Langmuir and circularly polarized transverse electromagnetic waves.
Note that the r.h.s. of \eref{eq:a} is proportional to
$n_e/\gamma_e-n_i/\gamma_i$ and thus represents the balance between
the striction nonlinearity due to perturbation of the density and
the relativistic nonlinearity due to mass. \Eref{eq:th} describes
the related phase evolution of the e. m. amplitude. For $a_0=0$, the
solution of \eref{eq:th} is simply $\theta=\theta_{0}-\bar k  \xi$,
so that the e. m. field  defined in (\ref{aperp}) has the simple
form $a(\xi) \exp[-i \bar \omega \tau +i \theta_{0}]$, being
$\tau=t-Vx$. For $a_0 \ne 0$, the phase evolution is non trivial.

The system (\ref{eq:phi}-\ref{eq:a}) can be put in Hamiltonian form, and has a first integral
\begin{equation}
H(a, a^{\prime}, \phi, \phi^{\prime})=
\frac{1-V^{2}}{2}\left( a^{\prime 2}+{\bar \omega}^{2 }a^{2}+{\bar k}^{2} \frac{a_{0}^{4}}{a^{2}} \right) -\frac{1}{2} \phi ^{\prime 2}-\gamma_{e}-\frac{\gamma_{i}}{\rho}
=K \label{eq:h}
\end{equation}
where the symbol $\prime$ denotes derivative with respect to $\xi$.
The value of the constant $K$ is determined by the boundary
condition.

For the case of a pure transverse electromagnetic wave (i.e., $\phi
=0$ and $a=a_{0}$), from \eref{eq:a}  the dispersion equation is
obtained \begin{equation} \omega ^{2}=\Omega^2+k^{2}, \qquad
\Omega^2=
\frac{1}{\sqrt{1+a_0^{2}}}+\frac{\rho}{\sqrt{1+\rho^{2}a_0^{2}}}
\label{disprel}
\end{equation}
$\Omega$ being the plasma frequency modified by the relativistic
effects. \Eref{disprel} corresponds to the Akhiezer-Polovin result
\cite{AP} with the ion motion taken into account.

So far, explicit reference to an electron--ion plasma has been made.
However, putting $\rho=1$, Eqs. (\ref{eq:phi}-\ref{eq:h}) are valid
also in a electron--positron plasma \cite{pre:ep}.

The particular class of localized solutions of equations
(\ref{eq:phi},\ref{eq:a})  is investigated, satisfying
$a^{\prime}=0$, $a^{\prime \prime}=0$, $\phi^{\prime}=0$, and
$\phi^{\prime \prime}=0$  at $\xi= \pm \infty$. For sake of
simplicity,  the case $\bar k =0$ is considered, so that $k= \omega
V$, and $\bar \omega = \omega$ (see e.g., \cite{KLS}). Note that
here $V$ is the group velocity.

\subsection{Quasineutral approximation}
Within the quasineutral approximation the main features of the
localized solutions can be identified in a simple manner, assuming
$n_{e}-n_{i}$, i.e.,  putting $\phi^{\prime \prime }=0$ \cite{KLS}.
Then, the electrostatic potential can be written in terms of the
vector potential amplitude, $\phi=
({\Gamma_{0i}\Gamma_{e}-\Gamma_{0e}\Gamma_{i}})/{\Gamma}$, being
$\Gamma=\Gamma_i+\rho \Gamma_e$, and the equation for the vector
potential in the quasineutral approximation becomes \cite{ppr01}
\begin{equation} a^{\prime \prime }+ {\bar \omega}^{2} a =a
\frac{V}{1-V^{2}} \frac{\Gamma^{2}} {\Gamma_{i} \Gamma_{e}
[\Gamma_{0}^{2}-(1-V^{2})\Gamma^{2}]^{1/2}}. \label{eq:qua-neu}
\end{equation}
The above equation has the following first integral
$H_{0}(a,a^{\prime})= {1}/{2}\left( a^{\prime 2}+{\bar \omega}^{2
}a^{2} \right)+{V \sqrt{\Gamma_{0}^{2}-(1-V^{2})\Gamma^{2}}}/{\rho
(1-V^{2})^{2}} =K_{0}  $,
and the solution for $a$ is reduced to quadrature. Then, in the
limit $\rho \ll 1$, the following simple expressions hold: $\phi
\approx \sqrt{1+a^2}-\sqrt{1+a_0^2}$, $n_{e}=n_{i}\approx 1+ \rho
\phi /V^2$, and  $v_{xe,i} \approx \rho \phi /V$.

From the analysis of the topology of the phase space $(a,a^{\prime
})$ of the Hamiltonian $H_0$,  different kind of solitary solutions
are found varying the propagation speed, and are summarized in
\tref{tbc}.
\begin{table}[h]
\caption{\label{tbc} Solitons in a cold plasma.} 
\begin{indented}
\item[]\begin{tabular}{@{}llllll} 
\br 
Propagation &Solution
&Potential $a$ &Potential $\phi$ &Density\\ speed &kind &at
$-\infty, +\infty$ & profile & profile\\ \mr $V^2>V_{s1}^2$ &bright
&$0$, $0$ &positive &accumulation\\ $V^2=V_{s1}^2$ &shock wave &$0$,
$a_1$ &positive &compression\\ $V^2_{s} < V^2 < V^2_c$ &dark
(gray) &$a_0$, $a_0$ &negative & evacuation\\ $V^2 = V^2_{s} $
&shock wave &$a_0$, $0$ &negative &rarefaction \\ $0< V^2 <V^2_{s}
$ &dark (black) &$a_0$, $-a_0$ &negative & evacuation\\ 
 \br
\end{tabular} 
\end{indented} 
\end{table}

Solitary solutions are found for $\omega^2 (1-V^2) <1+\rho$, with
dark solitons satisfying the dispersion relation (\ref{disprel}). In
both cases, the frequency spectrum  corresponds to the evanescent
region in the $(\omega, k)$ space, i.e., to the region $\omega^2 -k^2 < 1+\rho$. When the ion
dynamics is taken into account, bright solitons occurs for
propagation speed larger than a threshold value scaling with
$\sqrt{\rho}$, and dark solitons occur below this threshold value.
Standing bright solitons are found only when the ion dynamics is
neglected. Solutions in the form of
shock-waves are found at critical propagation speeds. These shock waves 
(occurring in the cold plasma limit) are collisionless, and similar to
those found at ion-acoustic speed \cite{mima:sw}.
The characteristic velocities  introduced in \tref{tbc} depend on
the potential amplitude and have the following expressions in the
low amplitude limit and for $\rho \ll 1$: $V_{s1}^2 \approx \rho
(1+a_1^2)$,  $V_{s}^2 \approx  \rho (1+a_0^4/16)$,  $V_c^2 \approx
\rho (1+a_0^2/2)$. In the same limit, bright solitons are described by 
$a=a_m \mathrm{sech}(\kappa \xi)$, with $\kappa^2=1+\rho-\omega^2 (1-V^2)$, and $a_m^2=4 \kappa^2/(1-\rho/V^2)$, while dark solitons by $a=a_0 \mathrm{tanh}(\kappa \xi)$, with $a_0^2 =4\kappa^2/(\rho/V^2-1)$, and $\omega^2 (1-V^2)=1+\rho-a_0^2/2$ \cite{KLS,ppr01}.

At large amplitude, quasineutrality is violated \cite{KLS}, and the
set of coupled equations (\ref{eq:phi},\ref{eq:a}) has to be solved
numerically. However, the above investigation on the existence of
the different kind of solutions has a general validity.

\subsection{Bright solitons in e--i plasma}

In the case of large amplitude bright solitons, a wide class of
solutions can be found. Here, the analysis is restricted to  the
class of solutions with single humped $\phi $ profiles, and $a$
profiles characterized by the number of nodes $p$. For fixed $V$,
the eigenspectrum of the bright solitons can been determined solving
numerically Eqs.  (\ref{eq:phi}-\ref{eq:a}) with suitable boundary
conditions \cite{KLS,prl01}.

\begin{figure}[ht] 
\begin{center} 
\includegraphics[width=2.1in]{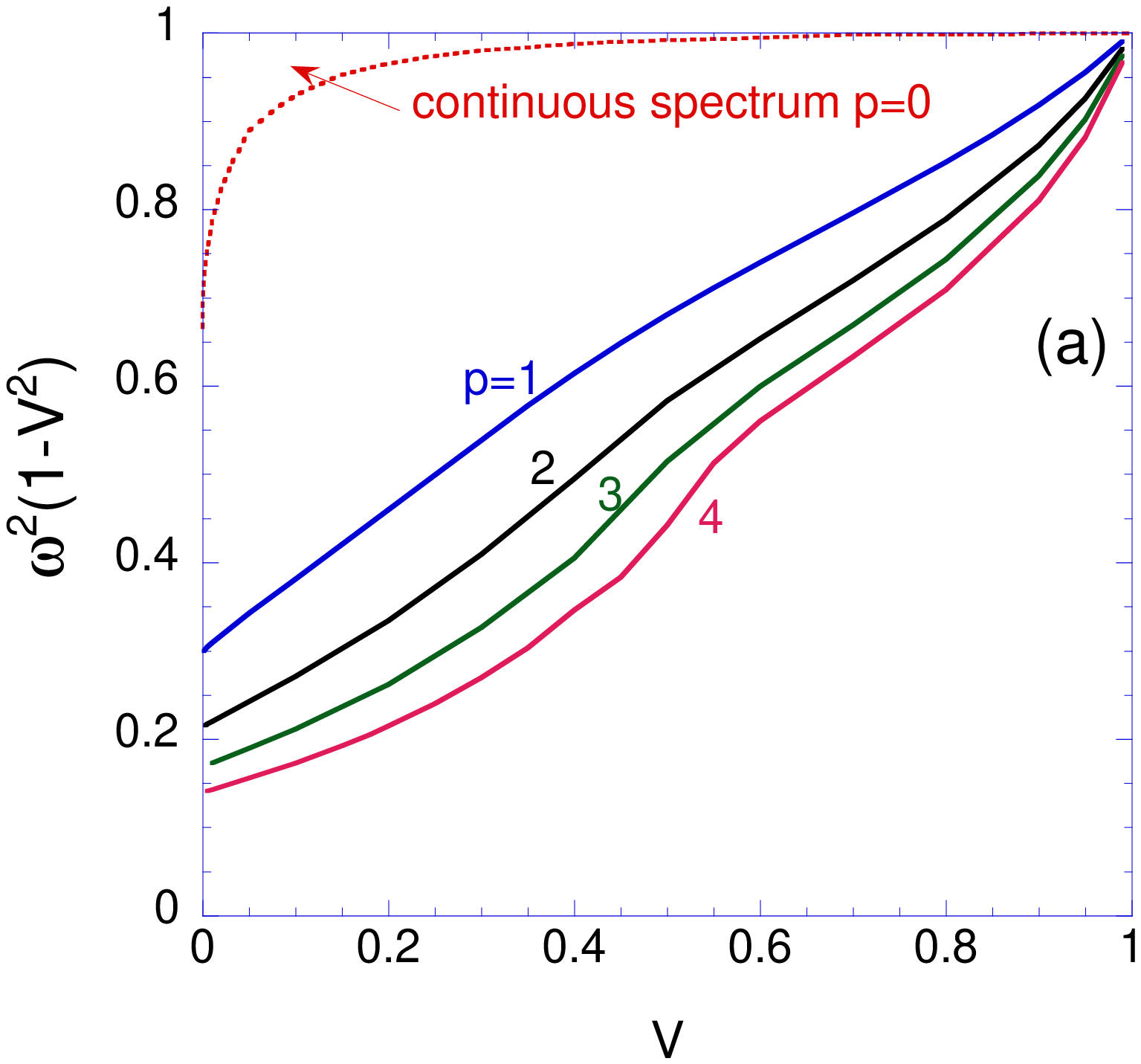}
\includegraphics[width=2.1in]{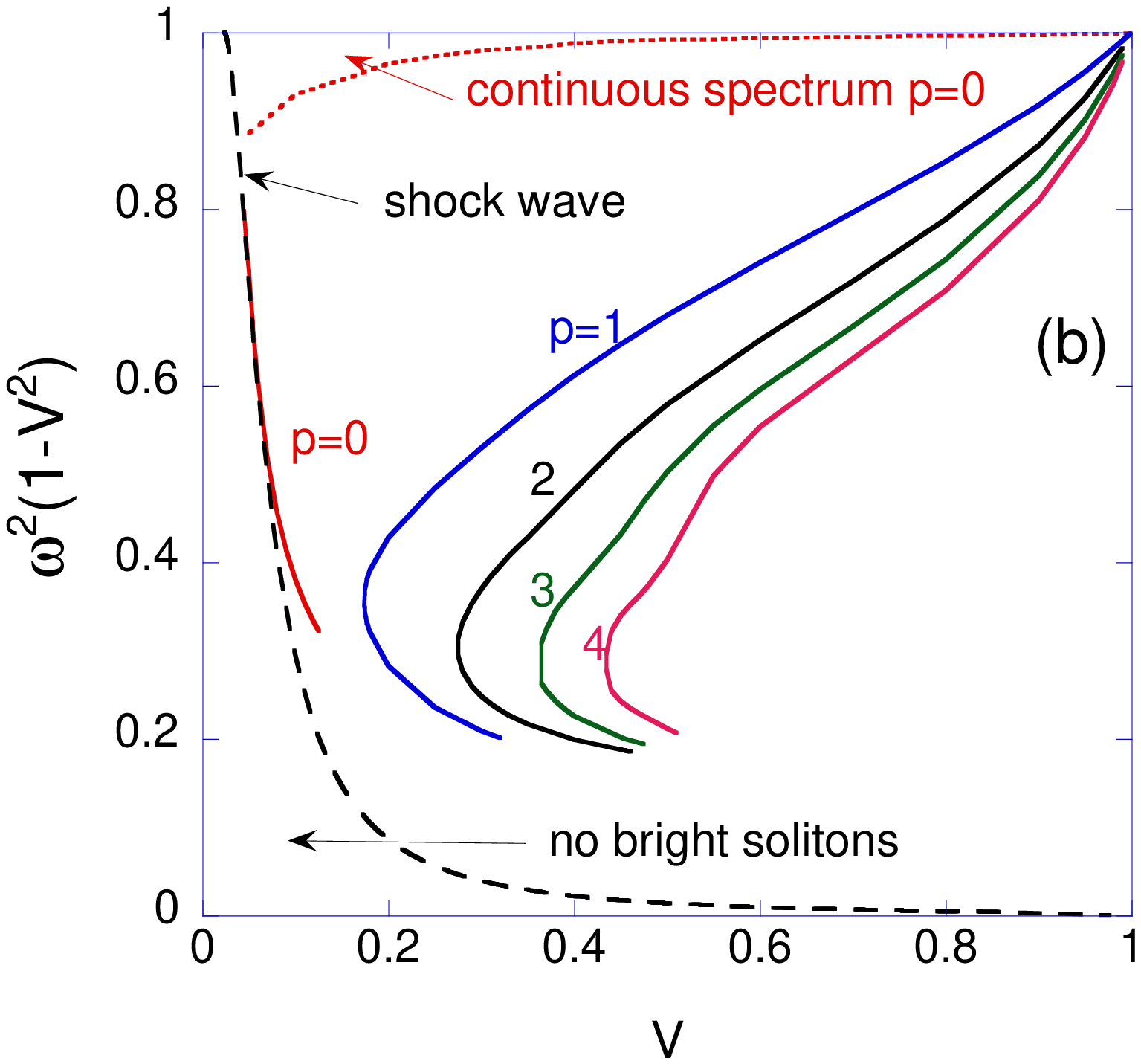}
\caption{Eigenspectrum of bright solitons as a function of the group
velocity $V$. Cases (a) and (b) refer to  $\rho=0$ (no ion dynamics)
and $\rho=1/1836$ (with ion dynamics), respectively. Both the
regions corresponding to the continuos spectrum of single--peaked
solitons and the discrete spectrum for $p=0,\ldots,4$ are shown. The
dashed curve in frame (b)  delimits a region of no solutions. }
\label{sp} 
\end{center} 
\end{figure}

The eigenspectrum relevant for low $p$ values is shown in \fref{sp},
both for the case in which the ions are fixed, i.e., at $\rho=0$,
and for the case in which the ion dynamics is taken into account.

When the ion dynamics is neglected (i.e.,  $\rho=0$), the bright
soliton eigenspectrum has the following features. Single-peaked
solitons ($p=0$) admit a continuous spectrum for any velocity value
\cite{poorn}. In particular, standing stable solitons ($V=0$) are
found with frequency  in the range $ 2/3 \le \omega^2 <1$, and $\max
{a} \le \sqrt{3}$, as it has been shown analytically in
\cite{Esirkepov:98}. Multi--peaked solitons $(p \ge 1)$ have a
discrete spectrum  for velocities larger than a small critical
value, and at fixed $V$, the frequency  $\omega$ decreases with
increasing $p$.  The peak value of the potentials $a$ and $\phi$ is
found at the minimum velocity, and increases with $p$ \cite{ppr01}.

\begin{figure}[ht]
\begin{center}
\includegraphics[width=1.5in]{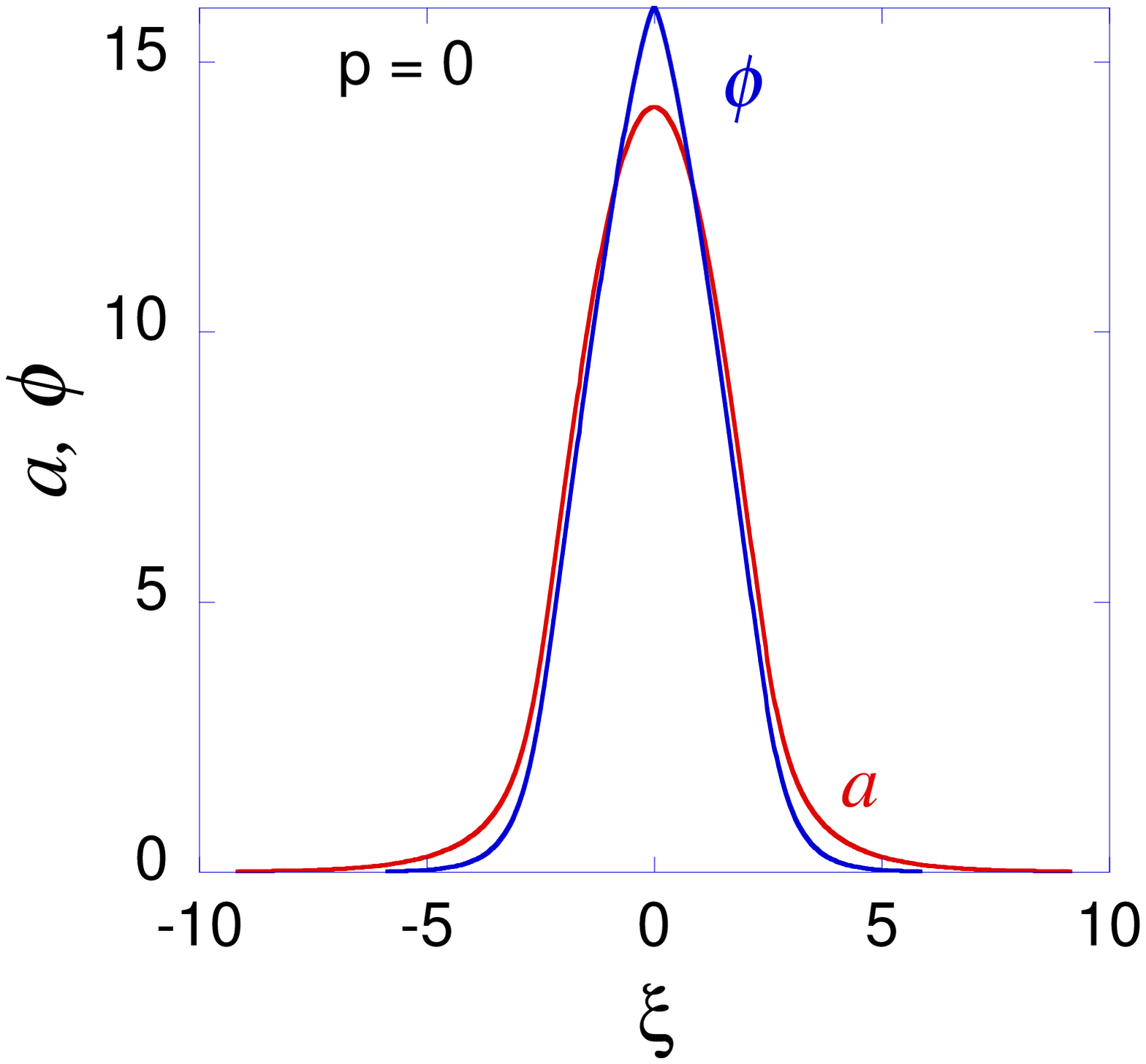}
\includegraphics[width=1.5in]{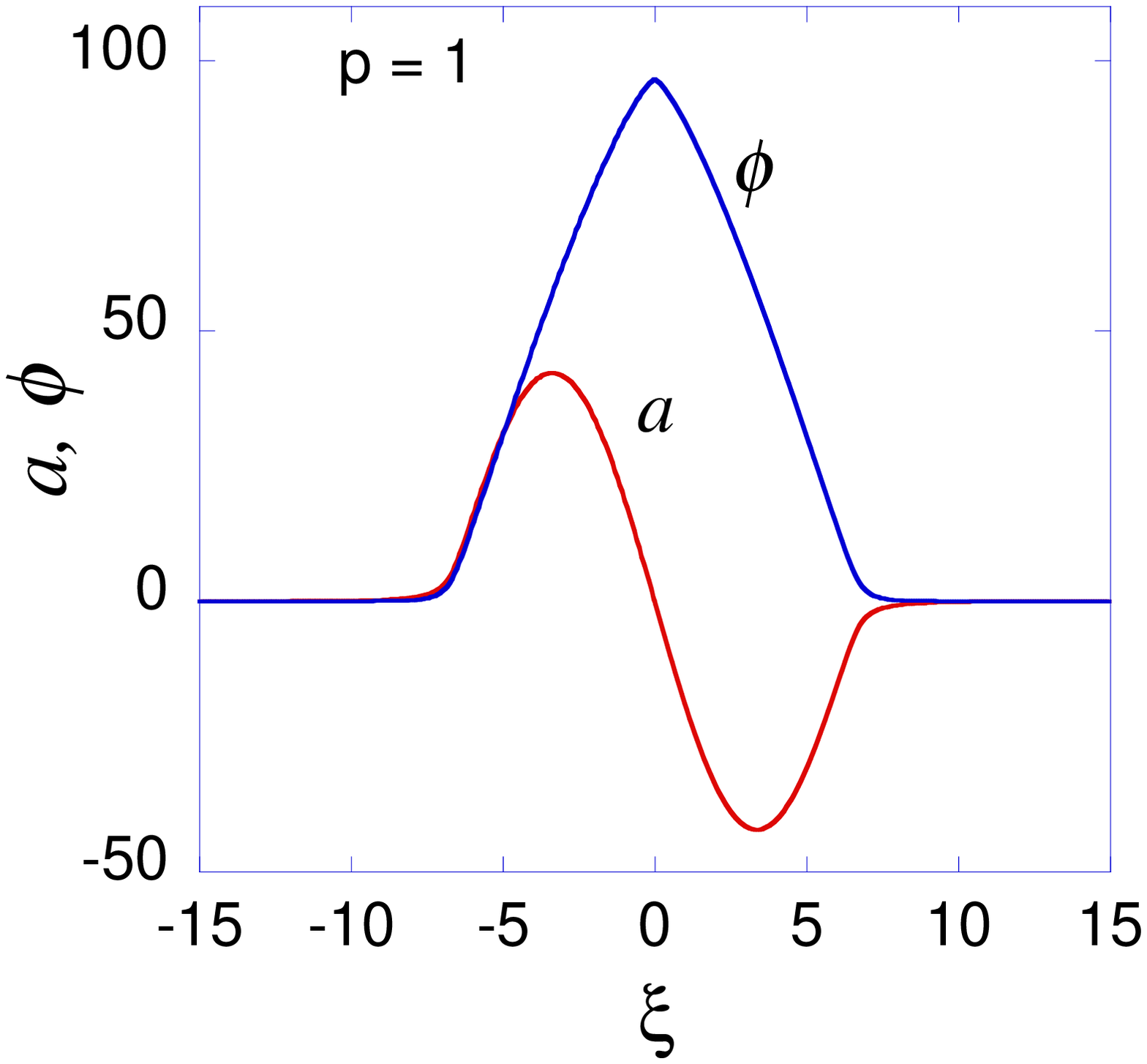}
\includegraphics[width=1.5in]{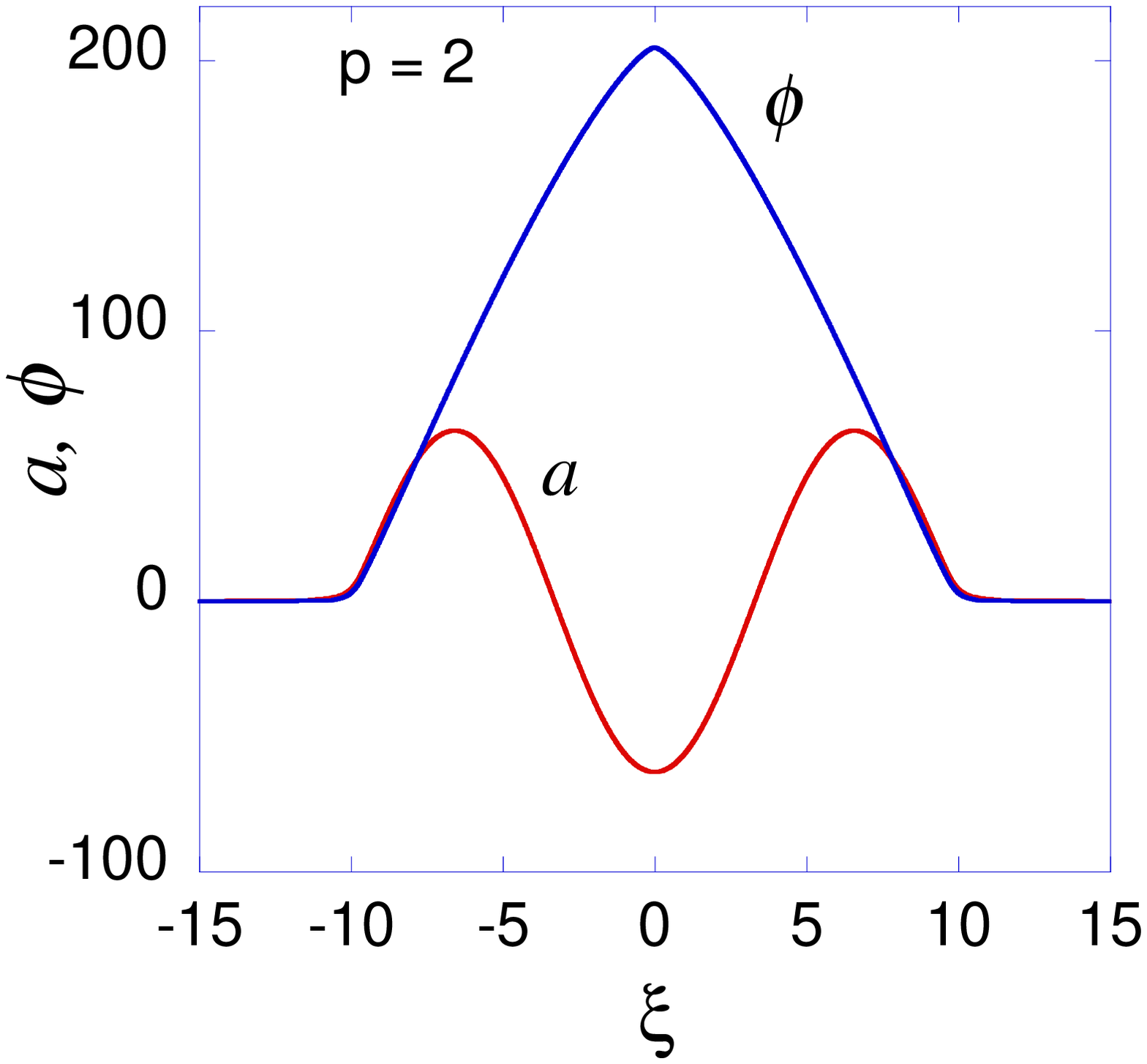}
\caption{Potential waveforms for bright solitons with $p=0,1,2$ and
velocities close to breaking.} \label{break}
\end{center}
\end{figure}

Quite different results are found when the ion dynamics is taken
into account (i.e., $\rho \ne 0$), even at a finite propagation
speed. Note that at high frequencies (and velocities), the ion
dynamics does not play any role. Single--humped solutions are found
for velocities larger than $V_s\approx \sqrt{\rho} $ \cite{KLS},
with both a continuous spectrum \cite{poorn} and a discrete spectrum 
\cite{prl01}. For $p \ge
1$, only a discrete spectrum is found \cite{prl01}. At low $p$
values, the frequency is non monotonous as a function of $V$, and at
$V=V_{br}$ the branch ends since the soliton breaks.   The structure
of the solution for $p=0,1,2$ and a velocity value close to breaking
is shown in  \fref{break}, where the potential waveforms are
plotted.   It is found that the ions  pile at the soliton center,
while the electrons at its edges, giving rise to very peaked density
distributions, and breaking occurs because the ion density diverges
in the center. At the same time, the ion and electron velocities
tend to $V_{br}$ at the center of the soliton and at its edges,
respectively, so that in these regions the particles move with
almost the group velocity of the soliton. The ion velocity profile
shows a cusp in the center, which is the signature of the nonlinear
wavebreaking.  The soliton breaking provides a novel mechanism for
the ion acceleration in the high intensity laser pulse interaction
with plasmas \cite{prl01}.

Solitons at large $p$ number  have been investigated numerically and
analytically in different approximations \cite{KSK,KZ,Dimant,poorn}.

\subsection{Dark solitons in e-i plasma}

\begin{figure}[ht]
\begin{center}
\includegraphics[width=2.1in]{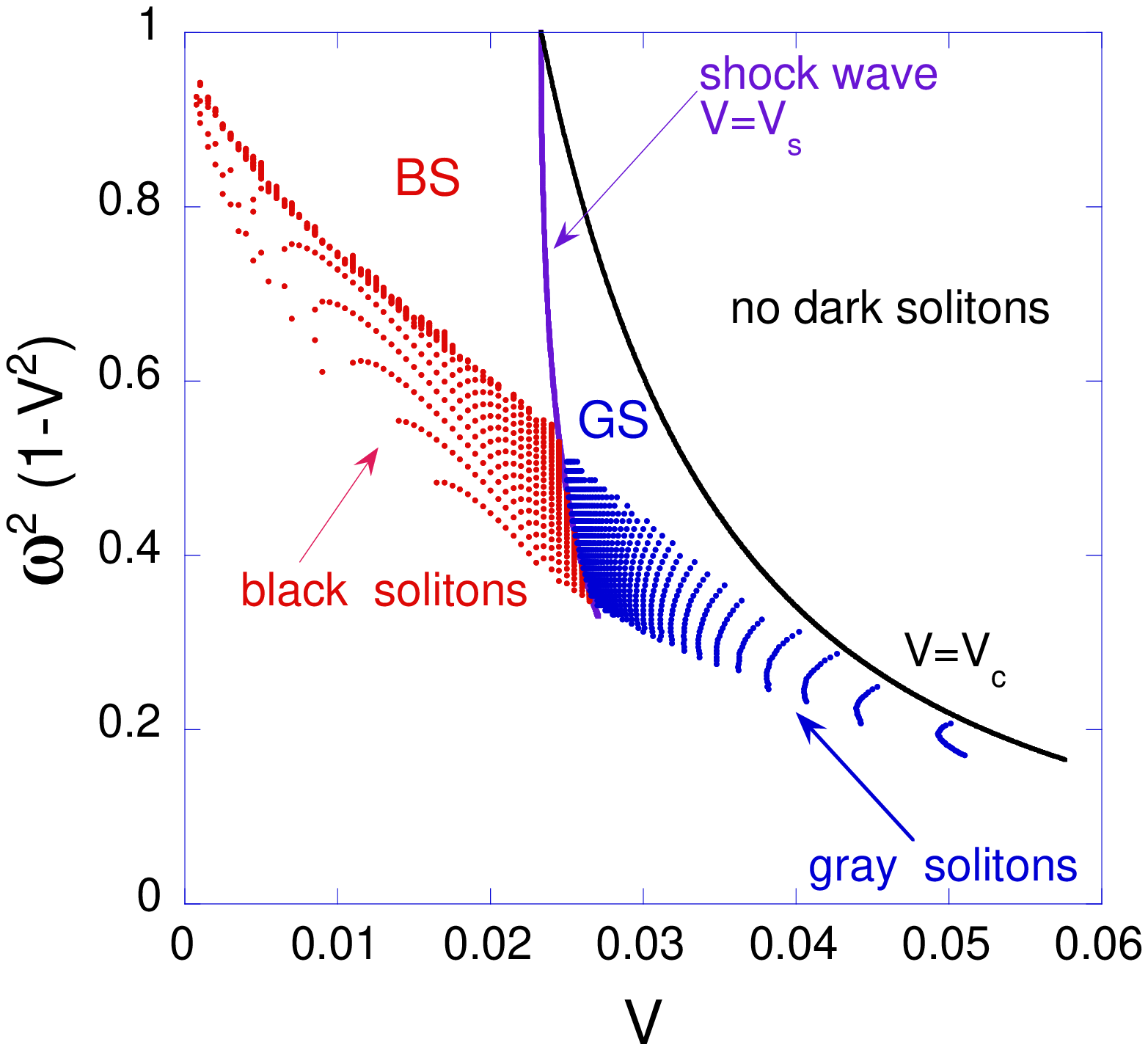}
\includegraphics[width=2.1in]{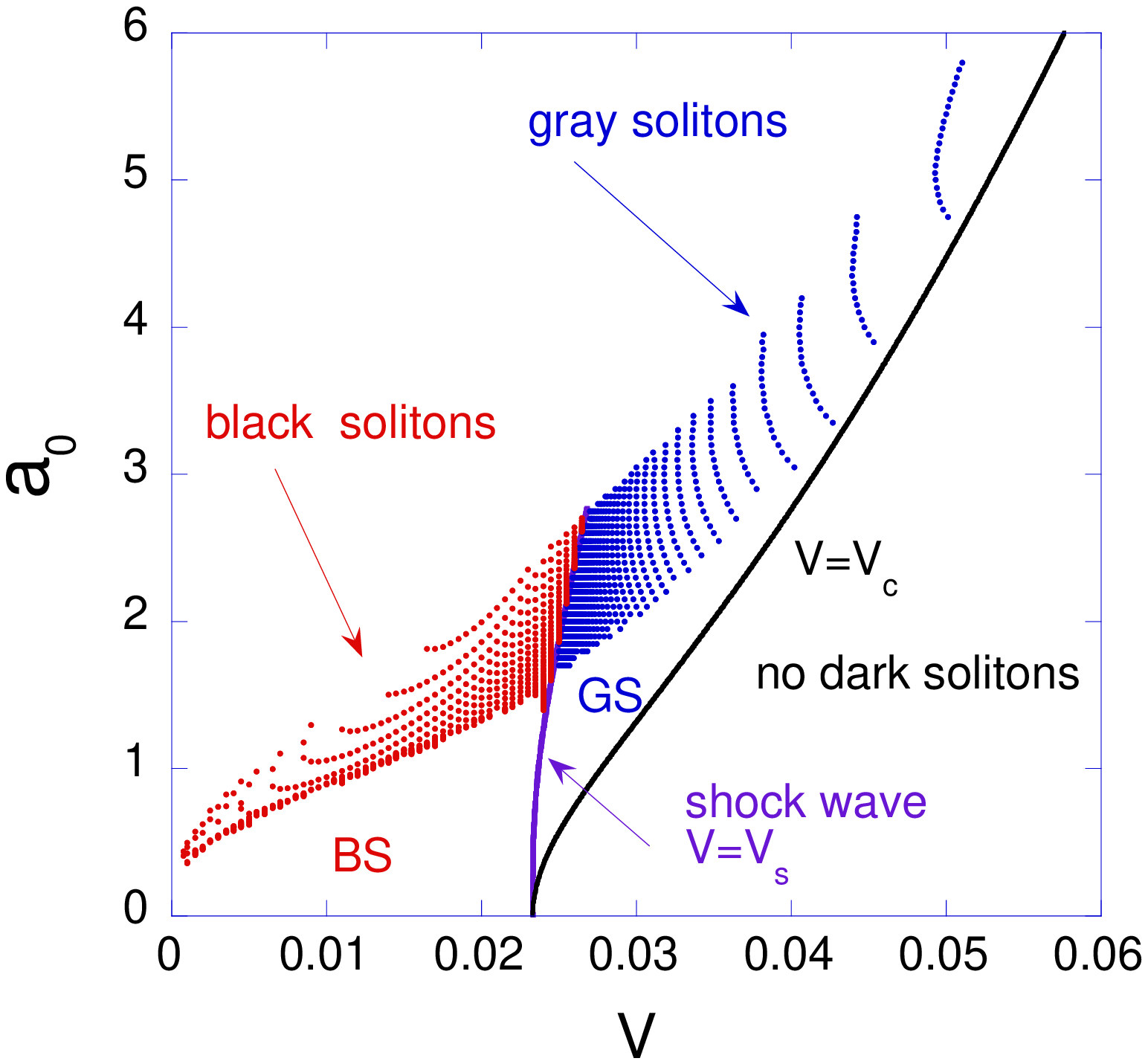}
\caption{Eigenspectrum of dark solitons (a) , and asymptotic e. m.
amplitude $a_0$ (b) as a function of the group velocity $V$. The
symbols GS and BS indicate the regions corresponding to the
continuous spectrum of gray and black dark solitons,  respectively.
The curve corresponding to shock waves is also plotted.} 
\label{ds}
\end{center} 
\end{figure}

\begin{figure}[ht] 
\begin{center}
\includegraphics[width=1.6in]{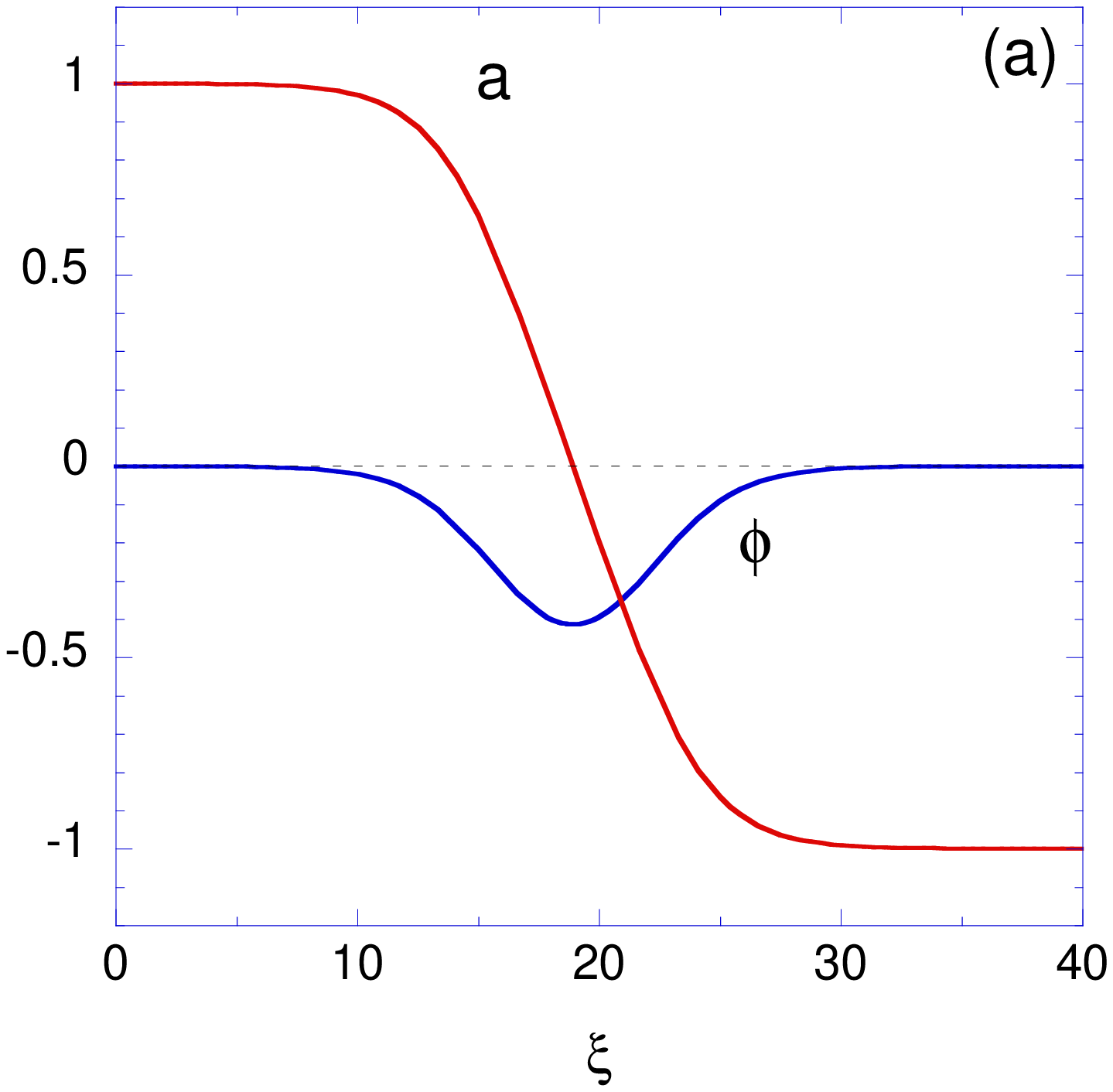}
\includegraphics[width=1.6in]{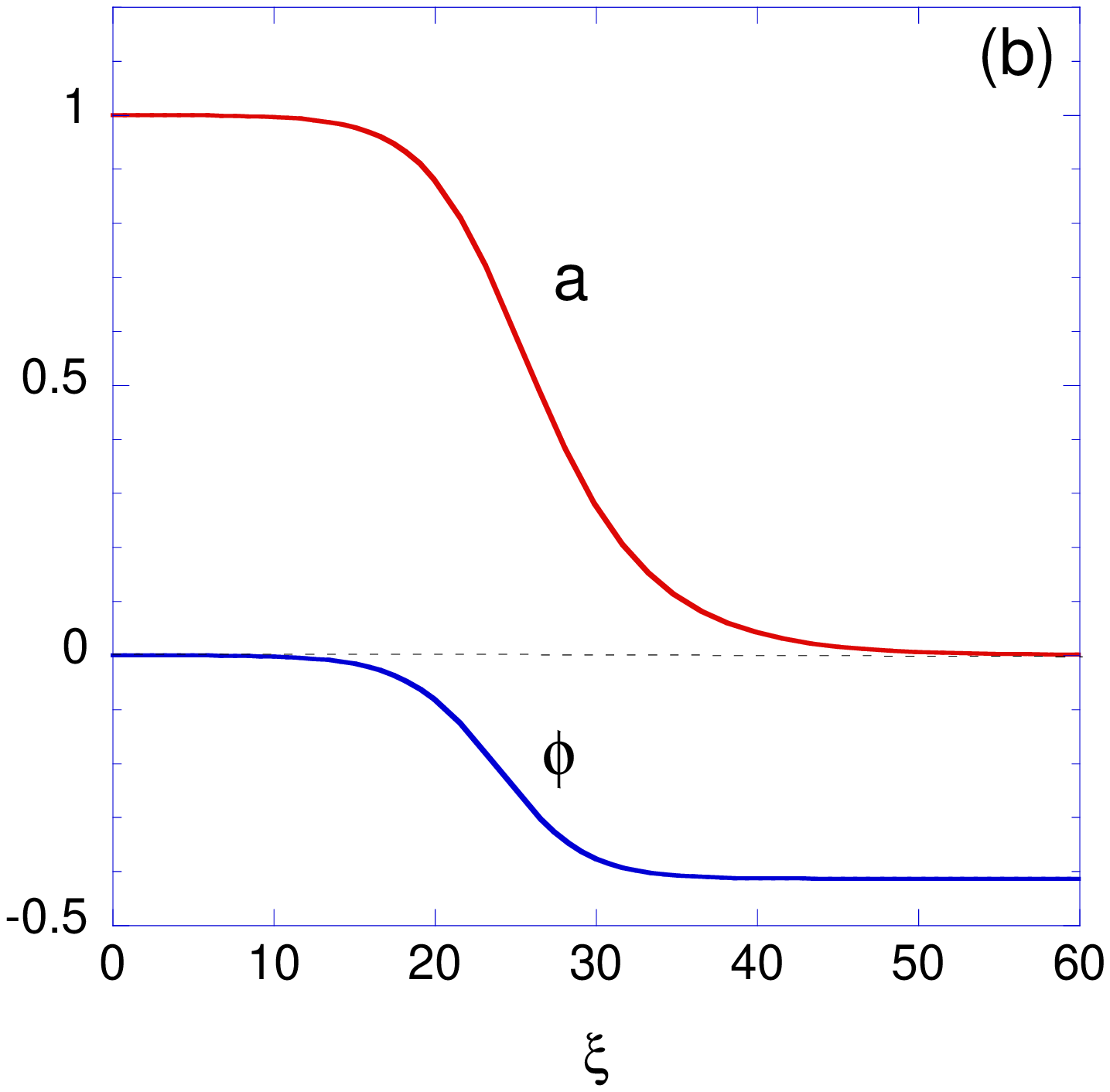}
\includegraphics[width=1.6in]{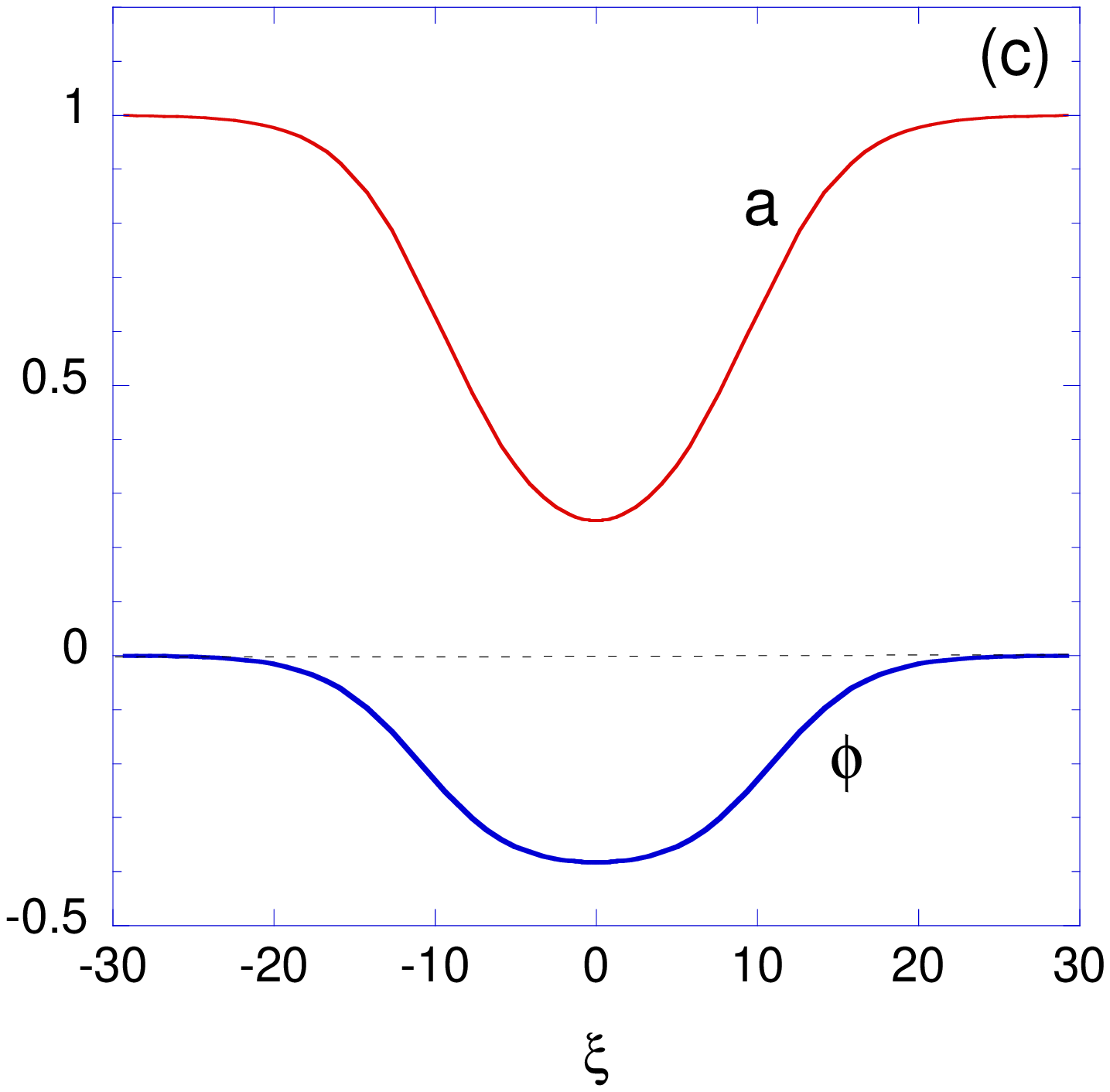}
\caption{Waveforms of a black soliton (a), a shock wave (b) and a gray
soliton (c). The chosen parameters are $a_0=1$, and $V<V_s$, $V=V_s$,
and $V_s<V<V_c$, respectively.} 
\label{wfds}
\end{center}
\end{figure}

As already pointed out, dark solitons are found only  when the ion
dynamics is taken into account and at very low velocities ($V
\lesssim \sqrt{\rho} $). The solutions of the full nonlinear system
for $a_0 \ne 0$ are obtained numerically, as in the case of bright
solitons \cite{aipproc}. The obtained spectrum and the corresponding
$a_{0}$ value are shown in \fref{ds} as a function of the velocity.
At fixed $V$, the spectrum is continuous at low $a_{0}$, and  it
becomes discrete above a threshold value, which increases with $V$.
The discrete spectrum is made up by a large number of separate
branches, each of them characterized by different densities and
velocities profiles. The ion density profile show always a dip,
while at large $a_0$ the electron density tends to peak in the
region inside the soliton. At a critical amplitude value (dependent
on $V$) breaking of the solutions occurs, due to the peaking of the
electron density. Solitary solutions are found up to $V \approx
0.051$, and $a_{0} \approx 5.8$ (for $\rho=1/1836$). Above these
values, no solutions have been found for the chosen parameters. Both
black and gray solitons are found for $V$ smaller and larger than
$V_s$, while a shock wave is found at $V=V_{s}$. \Fref{wfds} shows
the waveforms of the potentials for dark solitons (black and gray)
and the shock wave.

\subsection{Warm plasma effects}

The results described in the previous section have been derived in
the cold plasma approximation. However, it can be easily argued that
solitons moving  at ``low'' velocity can be affected by pressure
terms. The theoretical investigation of temperature effects has been
performed within the quasineutral approximation  for standing
solitons in a fully relativistic treatment \cite{mau}, and  for the
case of moving solitons in the weakly relativistic approximation
\cite{poorn:warm}. The results are summarized in \tref{tbw}
\cite{poorn:warm}, which refer to  small finite amplitude solitons
and non relativistic temperatures.
\begin{table}[h]
\caption{\label{tbw} Solitons properties in a warm plasma. The
following symbols are used: $V_{te,i}^2=T_{e,i}/m_{e,i}c^2$,
$c_s^2=\Gamma_e \rho V_{te}^2 +\Gamma_i V_{ti}^2$, with
$\Gamma_{e,i}$  the adiabaticity parameter.}
\begin{indented}
\item[]\begin{tabular}{@{}lllll}
\br
Propagation&Soliton&Potential $\phi$&Density &Species \\ speed&variety&at the center&at the center&involved\\ \mr $0\le V^2 <V_{ti}^2$&bright&positive&evacuation&i$+$e\\
$V_{ti}^2 < V^2 < c_{s}^2$&bright&negative&evacuation&i$+$e\\
$c_{s}^2 < V^2 < \rho +c_{s}^2$&dark&negative&evacuation&i$+$e\\
$\rho +c_{s}^2 < V^2 $&bright&positive&accumulation&i$+$e\\
$\rho +c_{s}^2 \ll V^2 $&bright&positive& evacuation&only e\\ \br
\end{tabular} \end{indented} \end{table}

As expected,  temperature effects play a crucial role in the range
$0 \le V^2< \rho +c_{s}^2 $, being $c_s$ the ion--acoustic speed.
Note that both inertia and temperature terms drive  the process of
soliton formation in a warm plasma. In particular, the formation of
single-peaked solitons is observed  at zero and very low propagation
speed, lower than the ion acoustic speed. The relevant electron and
ion density profiles show a dip in the center of the soliton. In the
low temperature limit, density depletion in the center of standing
solitons may become almost total \cite{mau}.

\section{Solitons in laser plasma simulations}

Relativistic solitons has been investigated with the help of fluid
simulations and  2D, 3D PIC simulations for laser--plasma
interaction \cite{sim92,sim95,sim99, postsol,sim02,sim04}. The mechanism
of soliton generation by a high intensity laser pulse propagating in
a underdense plasma can be descibed as follows. The laser pulse
interacting with the plasma loses its energy,  as it generates wake
fields behind itself. Since the process is adiabatic the ratio
between the e. m. energy density and the frequency is conserved, so
that frequency downshift occurs for the wake field.  Moreover, since
the group velocity of the laser pulse decreases with the frequency,
the low frequency part of the e. m. radiation of the pulse
propagates with very low velocity. When the frequency becomes lower
than the Langmuir frequency, the e. m. energy becomes trapped inside
the related density cavity, thus forming a slowly moving solitary
structure. Note that in 2D and 3D it is found that solitary waves
can merge, thus they are not strictly speaking solitons. It has been
observed  that a large part of the laser pulse energy can be
transformed into solitons \cite{sim99}. It can then be argued that
solitary waves may play an important role in the laser--plasma
interaction.

When PIC simulations are performed in inhomogeneous plasmas
\cite{Sentoku-99},  it is found that low frequency solitary waves
generated by superintense laser pulses are accelerated along the
density gradient towards the lowest density values (e.g., the
plasma--vacuum interface) where they radiate their energy in the
form of low frequency electromagnetic bursts. This process can be a
signature of soliton formation.

PIC simulations show that the time  of the soliton formation is much
shorter than the ion response time, so that ions can be assumed at
rest during this process. For approximately $ (m_i/m_e)^{1/2} $
oscillation periods of the e. m. field inside the soliton (which is
of the order of $2 \pi/\omega_{pe}$), the ions can be assumed as
fixed. For longer times, the ponderomotive force starts to dig a
hole in the ion density, and the parameters of the solitons change.
What was a soliton on the electron timescale is no more a soliton on
the ion timescale. Nevertheless, a low  frequency  e. m. wave packet
remains well confined inside a slowly expanding plasma cavity. This
e. m. entity has been called a postsoliton \cite{postsol}.

In addition, 1D fluid simulations in ``slightly''  overdense plasmas
(i.e., for plasma densities $\lesssim 1.5$ times the critical
density)  have shown that penetration of relativistically intense
laser radiation occurs by soliton--like structures moving into the
plasma \cite{tush}.

\section{Final remarks}

Many theoretical problems are still open even for 1D solitons, as,
e.g., the temperature effects at any propagation speed and arbitrary
wave amplitude, the soliton dynamics in the presence of plasma
inhomogeneity,  the properties of linearly polarized solitons, and
last but not least the stability of such structures (e. g., see
discussions in Ref. \cite{jovo}).

In conclusion, we point out that  the first macroscopic evidence of
soliton formation in the interaction of an intense ($10^{19}$
W/cm$^2$) laser pulse with an underdense plasma has been reported in
\cite{borghesi}. Long-lived, macroscopic bubble--like structures
have been detected through the deflection that the associated
electric charge separation causes in a proton probe beam
\cite{prima}. These structures are interpreted as the remnants of a
cloud of relativistic solitons generated in the plasma by the
ultraintense laser pulse.

\end{document}